\documentclass[aps,prl,twocolumn]{revtex4}
\usepackage{graphicx}
\usepackage{amssymb}
\usepackage[dvips]{color}
\newcommand{\bfr}{{\bf r}}
\newcommand{\bfq}{{\bf q}}

\newcommand{\um}{{U_m}}
\newcommand{\hpsi}{{\hat\psi}}
\newcommand{\hPsi}{{\hat\Psi}}
\newcommand{\hphi}{{\hat\phi}}
\newcommand{\ha}{{\hat a}}
\newcommand{\omz}{{\omega_0}}
\newcommand{\lz}{{l_0}}

\begin{document}
\title{Confinement effects on the stimulated dissociation of molecular BECs}
\author{Igor Tikhonenkov and Amichay Vardi}
\affiliation{Department of Chemistry, Ben-Gurion University of the
Negev, P.O.B. 653, Beer-Sheva 84105, Israel} 

\begin{abstract}
We show that a molecular BEC in a trap is stabilized against stimulated dissociation if the trap size is smaller than the resonance healing length $(\hbar^2/2mg\sqrt{n})^{1/2}$. The condensate shape determines the critical atom-molecule coupling frequency. We discuss an experiment for triggering dissociation by a sudden change of coupling or trap parameters. This effect demonstrates one of the unique collective features of 'superchemistry' in that the yield of a chemical reaction depends critically on the size and shape of the reaction vessel. 
\end{abstract}


\maketitle

Atom-molecule conversion in quantum bose gases is equivalent to three-wave mixing in nonlinear optics \cite{Drummond98,Javanainen99,Kostrun99,Vardi01,Meystre05,Rowen05}. Association is analogous to second-harmonic generation and dissociatiocorresponds to parametric downconversion. The latter process exhibits a modulational instability, resulting in from exponential amplification of spontaneously emitted photon-pairs \cite{YarivWalls}. The existence of similar parametric gain in molecular dissociation was noted early on in the study of atom-molecule condensates \cite{Kostrun99,Vardi01,superchem,Moore02}. This dynamical instability generates significant deviations from the Gross-Pitaevskii (GP) classical field theory \cite{Vardi01,quantpa}. The rapid growth of correlations can be used for the generation of pair-correlated or number-squeezed atomic beams \cite{Kheruntsyan} in analogy to the use of parametric downconversion to generate squeezed light \cite{WallsYurke}. 

Exponential gain in molecular BEC dissociation is readily captured within an undepleted pump approximation \cite{Vardi01,Moore02,Kheruntsyan}. The atom-molecule Hamiltonian is,
\begin{eqnarray}
H&=&\int d\bfr\left\{\hpsi^\dag(\bfr)\left[-\frac{\hbar^2}{2m}\nabla^2+V(\bfr)+\Delta\right]\hpsi(\bfr)\right.\nonumber\\
~&~&+\hphi^\dag(\bfr)\left[-\frac{\hbar^2}{4m}\nabla^2+2V(\bfr)+\frac{\um}{2}\hphi^\dag(\bfr)\hphi(\bfr)\right]\hphi(\bfr)\nonumber\\
~&~&+\left.\frac{g}{2}\left[\hphi^\dag(\bfr)\hpsi(\bfr)\hpsi(\bfr)+\hpsi^\dag(\bfr)\hpsi^\dag(\bfr)\hphi(\bfr)\right]\right\}~,
\label{ham}
\end{eqnarray}
where $\hpsi$ and $\hphi$ are atomic and molecular bose field operators, $m$ is the atomic mass, $\Delta$ is the detuning from atom-molecule resonance and $V(\bfr)=m\omz^2 r^2/2$ is an isotropic magnetic or optical trap potential with frequency $\omz$. Atoms and molecules see essentially the same trap frequency since molecules have twice the mass and twice the magnetic moment or optical polarizability \cite{Greiner03}. Atom-molecule coupling of strength $g$, is achieved by a Feshbach resonance \cite{Tiesinga93} or by an optical stimulated Raman transition \cite{Band95}. The molecule-molecule interaction strength is $\um$ and interactions between atoms are neglected since we are interested in the initial dissociation dynamics when atomic densities are small. The Heisenberg equations of motion for the field operators read,
\begin{eqnarray}
\label{psidot}
i\hbar\frac{\partial}{\partial t}\hpsi&=&\left[-\frac{\hbar^2}{2m}\nabla^2+\Delta+V(\bfr)\right]\hpsi+g\hpsi^\dag\hphi~,\\
~&~&~\nonumber\\
\label{phidot}
i\hbar\frac{\partial}{\partial t}\hphi&=&\left[-\frac{\hbar^2}{4m}\nabla^2+2V(\bfr)+\um\hphi^\dag\hphi\right]\hphi+\frac{g}{2}\hpsi^2~.
\end{eqnarray}
The undepleted pump approximation is a linearization of Eq. (\ref{psidot}) about a stationary, classical molecular field $\phi(\bfr)$ obeying the GP equation 
\begin{equation}
\label{phiGP} 
-\frac{\hbar^2}{4m}\nabla^2\phi+[2V(\bfr)+\um|\phi|^2]\phi=\mu\phi~.
\end{equation}
Replacing $\hphi(\bfr,t)\rightarrow\phi(\bfr)e^{-i\mu t/\hbar}$, and rotating $\hPsi=\hpsi e^{i\mu t/ 2\hbar}$, $\hPsi^\dag=\hpsi^\dag e^{-i\mu t/ 2\hbar}$ (so that the atomic density remains $n_a=\langle\hpsi^\dag\hpsi\rangle=\langle\hPsi^\dag\hPsi\rangle$), we obtain the coupled linear equations for the atomic field operators $\hPsi$ and $\hPsi^\dag$,
\begin{eqnarray}
\label{stabo}
i\hbar\frac{\partial}{\partial t}\hPsi&=&\left[-\frac{\hbar^2}{2m}\nabla^2+\delta+V(\bfr)\right]\hPsi+g\phi\hPsi^\dag,\\
\label{stabt}
-i\hbar\frac{\partial}{\partial t}\hPsi^\dag&=&\left[-\frac{\hbar^2}{2m}\nabla^2+\delta+V(\bfr)\right]\hPsi^\dag+g\phi^*\hPsi,
\end{eqnarray}
where $\delta=\Delta-\mu/2$ is an effective detuning including the molecular chemical potential consisting of zero-point energy and of a molecular interaction shift. The molecular condensate stability is determined by the eigenvalues $\lambda$ of the set (\ref{stabo})-(\ref{stabt}). For a uniform gas with $V(\bfr)=0$, $\phi=\sqrt{n}$, and $\mu=\um n$ ($n$ being the molecular density), the system is diagonalized in momentum space $\hpsi(\bfr,t)=\sum_\bfq \ha_\bfq(t)\exp(i\bfq\cdot\bfr)$, to give the dispersion,
\begin{equation}
\label{lambdaq}
\hbar\lambda_q=\sqrt{(\epsilon_q+\delta)^2-ng^2},
\end{equation}
where $\epsilon_\bfq=(\hbar q)^2/2m$ is the free-particle dispersion. Thus for a uniform gas with $\delta<\sqrt{n}|g|$, there are unstable resonant modes for which $\epsilon_\bfq+\delta<g\sqrt{n}$, resulting in complex characteristic frequencies and exponential gain. The dynamics of amplified modes is described by well-known solutions, given in terms of the hyperbolic functions $\cosh(\lambda_q t)$ and $\sinh(\lambda_q t)$, which can be found in Refs. \cite{Vardi01,Kheruntsyan,WallsYurke}. For $\delta=0$, the unstable modes are simply low-energy, long-wavelength excitations.

To gain insight on nonuniform molecular BEC stability, we compare Eqs.(\ref{stabo})-(\ref{stabt}) to the standard Bogoliubov analysis \cite{Pethick} of an attractively interacting BEC, with the uniform-gas dispersion $\lambda_q=\sqrt{\epsilon_q(\epsilon_q+2nU)}$, where $U$ is the interaction strength. The difference between the Bogoliubov dispersion relation and  Eq. (\ref{lambdaq}) results from the existence of additional mean-field diagonal terms in the Bogoliubov equations, corresponding to different pairings of the $U\hpsi^\dag\hpsi\hpsi$ interaction term. For attractive interaction $U<0$, the long-wavelength modes with $\epsilon_q<|2nU|$ are unstable, leading to BEC collapse \cite{Collapse}. For a BEC confined to a trap however, collapse is prevented provided that the size of the trap is short with respect to the healing length $\xi=\sqrt{\hbar^2/(2mnU)}$. Since the unstable modes are those excitations with wavelength {\it longer} than $\xi$, they are precluded from tighter traps. Therefore, stable condensates of attractively interacting atoms exist up to a critical density for a given trap-frequency \cite{Attractive}.

 A similar instability threshold exists for molecular BEC dissociation. Even for $\delta=0$, there will be no unstable modes towards dissociation provided that the trap size $\lz=\sqrt{\hbar/(m\omz)}$ is smaller than the 'resonance healing length' $\zeta=\sqrt{\hbar^2/(m gn^{1/2})}$, which is the characteristic lengthscale of Eqs. (\ref{stabo}) and (\ref{stabt}), resulting from balancing  kinetic and resonance energies. Thus for sufficiently tight confinement, atomic zero-point motion inhibits dissociation. Remarkably, the outcome of a coherent chemical process depends due to its collective nature, not only on the statistics of the constituent atoms, but also on the size and geometry of the 'container' in which it is carried out. 
\begin{figure}
\centering
\includegraphics[scale=0.4,angle=0]{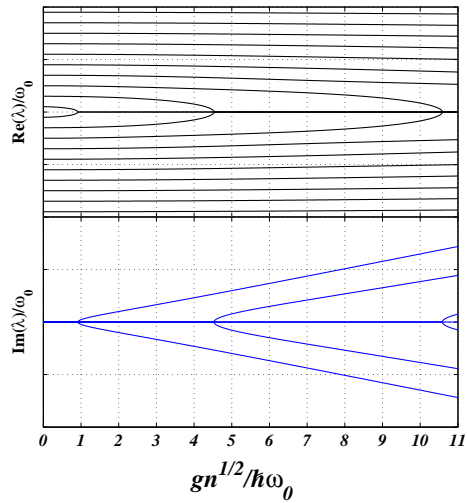}
\caption{(color online) Real and imaginary parts of characteristic dissociation frequencies as a function of atom-molecule coupling strength, for a harmonic trap with frequency $\omz$. The coupling is assumed to be resonant, i.e. $\delta=0$ and the molecular condensate is noninteracting, i.e. $U_m=0$ . Complex frequencies only appear when the coupling strength $g\sqrt{n}$ is larger than the trap level spacing $\hbar\omz$}
\label{bosdis_fig1}
\end{figure}

In order to ascertain that confinement can stabilize a molecular BEC against stimulated dissociation, we have calculated the characteristic frequencies of Eqs. (\ref{stabo})-(\ref{stabt}) with a finite-size undepleted pump $\phi$ obtained from Eq. (\ref{phiGP}), by expanding 
$\phi(\bfr)=\sum_j \varphi_j \chi_j(\bfr)$, $\hPsi(\bfr,t)=\sum_j \ha_j(t)\chi_j(\bfr)$ in the basis of harmonic-oscillator solutions $\chi_j(\bfr)$, and diagonalizing the resulting linear set of equations for the dynamics of $\ha_j(t)$ and $\ha_j^\dag(t)$,
\begin{equation}
i\frac{d}{dt}\left(\begin{array}{c}
{\bf {\hat a}}\\
{\bf {\hat a}^\dag}
\end{array}\right)=
\left(\begin{array}{cc}
\Omega&\Phi\\
-\Phi^\dag&-\Omega
\end{array}\right)
\left(\begin{array}{c}
{\bf {\hat a}}\\
{\bf {\hat a}^\dag}
\end{array}\right),
\end{equation}
with ${\bf {\hat a}}=(\ha_0,\ha_1,...,\ha_j,...)$, ${\bf {\hat a}^\dag}=(\ha^\dag_0,\ha^\dag_1,...,\ha^\dag_j,...)$, $\Omega_{ij}=\omz\delta_{ij} (j+1/2)$, $\Phi_{ij}=(g/\hbar)\sum_l \varphi_l\langle\chi_j\chi_l|\chi_i\rangle$. The results were compared to a discrete Fourier transform diagonalization, giving precisely the same eigenfrequencies. Eigenvalues of a one-dimensional calculation with $\delta=U_m=0$, are plotted in Fig. \ref{bosdis_fig1} as a function of the atom-molecule conversion frequency $g\sqrt{n}/\hbar$. Complex frequencies, corresponding to an initially exponential molecular gain, are only obtained if the coupling frequency is larger than the trap frequency. Thus, for $g\sqrt{n}/\hbar\omz<\alpha$ where $\alpha$ is a geometrical factor of order one, the molecular condensate is stabilized by atomic zero point motion. 

\begin{figure}
\centering
\includegraphics[scale=0.4,angle=0]{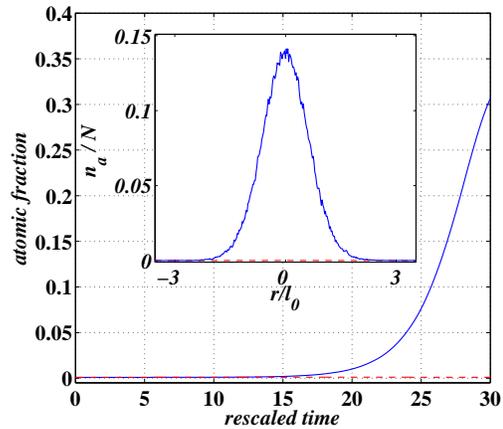}
\caption{(color online) Total atomic fraction obtained from a mean-field calculation, triggered by Gaussian noise, as a function of rescaled time $\tau=\omz t$ for $g\sqrt{n}=0.9\hbar\omz$ (dashed red line) and for $g\sqrt{n}=\hbar\omz t$ (solid blue line), with $\delta=U_m=0$. The corresponding atomic density profiles at the final time $\tau=30$ are shown in the inset.}
\label{bosdis_fig2}
\end{figure}

In Fig. \ref{bosdis_fig2} we plot the results of an approximate solution of the dynamical equations (\ref{psidot})-(\ref{phidot}), including pump depletion. The equations were solved in a mean-field approach, simulating initial spontaneous dissociation by a complex Gaussian white noise with zero mean, corresponding to atomic quantum fluctuations in the Wigner representation \cite{Santagiustina98,Gatti97}. While this approach does not fully capture the quantum dynamics in the way full stochastic calculations do \cite{Kheruntsyan}, it  suffices to depict the dominant quantum effect of amplification of spontaneously emitted atom pairs in the vicinity of the instability \cite{Vardi01,Moore02}. Since we are interested here in the mere presence of an instability, our calculation should indicate whether molecular gain exists even if it does fail to give the precise onset time. Evidently, Fig. \ref{bosdis_fig2} demonstrates the existence of an amplification threshold $g_c\approx 0.91\hbar\omega/\sqrt{n}$. For $g<g_c$ all frequencies are real and there is no observed dynamical gain, whereas for $g$ just above the critical value, molecular amplification takes place.

\begin{figure}
\centering
\includegraphics[scale=0.4,angle=0]{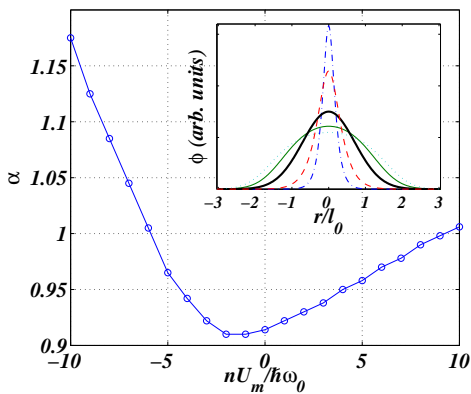}
\caption{(color online) Critical coupling strength $\sqrt{n}g_c$ as a function of the molecule-molecule interaction frequency $nU_m$. Magnetic or optical atom-molecule detunings $\Delta$ were assumed to compensate the molecular chemical potential, so that $\delta=0$ for each interaction frequency. In this way, true shape effects are distinguished from mean-field shifts originating from the variation in the molecular chemical potential $\mu$.}
\label{bosdis_fig3}
\end{figure}

The precise critical ratio between coupling strength and trap frequency depends upon the shape of the molecular condensate. In Fig \ref{bosdis_fig3} we plot the numerically-obtained value of $\alpha$ as a function of the molecule-molecule interaction strength $nU_m$, in a one-dimensional calculation. The condensate profile becomes broader with increasing repulsive interaction (see inset), leading to higher values of $g_c$ for a given $\omz$ and $n$. Since the calculation is in 1D, stable molecular condensates with self-trapped solitary profiles, significantly narrower than the trap size, do exist also for attractive interaction. The lowest value of $\alpha$ is obtained at the transition to self-trapping at $nU_m\approx-\hbar\omz$. 

The coupling energy scales as $g\sqrt{n}=g(N/\Lambda_d\lz^d)^{1/2}$ where $d$ is the number of dimensions and $\Lambda_d$ is a $d$-dependent geometrical factor. Since the spacing between trap levels is $\hbar\omz=\hbar^2/m\lz^2$, we obtain that at the critical point $\sqrt{N/\Lambda_d}\lz^{2-d/2}\approx\hbar^2/(mg)$. Thus for a three-dimensional system $d=3$, $\Lambda_3\approx 8\pi \sqrt{2}/3$ and a fixed trap size, the critical number of particles in the molecular condensates is $N_C\approx \Lambda_3\left(\frac{\hbar^2}{mg}\right)^2/\lz$. This critical number should be compared with the critical number of particles for the collapse of a 3D BEC with attractive interaction of strength $U=4\pi\hbar^2 a_s/m<0$ ($a_s$ being the atomic $s$-wave scattering length), obtained from balancing $2Un=2UN/(\Lambda_3\lz^3)$ with $\hbar\omz=\hbar^2/m\lz^2$, to give $N_C\approx \Lambda_3\hbar^2\lz/(2m|U|)=\frac{\sqrt{2}}{3}\lz/a_s$. Thus, in contrast to the critical number of particles for collapse, larger molecular condensates will be stabilized against stimulated dissociation for tighter traps, because the increase in zero point motion overcomes the effect of increasing density. 

Most atom-molecule systems coupled by a magnetically tunable Feshbach resonance, are well within the strong-coupling, nonlinear domain, where the characteristic atom-molecule conversion frequency is much larger than the trap frequency. For example for the $^{23}$Na Feshbach resonance at 907 G \cite{Inouye98,Stenger99}, the coupling strength is given by $g=\sqrt{\kappa U_a}$ \cite{Timmermans99,Holland01} with $U_a=4\pi\hbar^2a_s/m$, $\kappa/(2\pi\hbar)\approx4.6$ MHz, and $a_s\approx60$ Bohr. These values give $g/2\pi\hbar\approx 9\times10^{-3}$ Hz cm$^{3/2}$. Given a density of $n=10^{15}$ cm$^{-3}$, we have $g\sqrt{n}/2\pi\hbar\approx 284$ kHz, far above contemporary trap frequencies (we note that comparison of this Feshbach frequency to the atomic interaction frequency $U_a n/2\pi\hbar\approx 17.5$ kHz, justifies the neglect of atom-atom interactions in our calculations, even for substantial atomic populations and high densities. For $n=10^{13}$cm$^{-3}$, $g\sqrt{n}/2\pi\hbar$ drops to $28.4$ kHz, whereas $U_a n/2\pi\hbar$ is only $175$ Hz). For $\omz/2\pi=200 Hz$ we obtain that $N_c\approx 2\times10^{-2}$, so that stimulated dissociation will be  observed for any number of molecules in the condensate. It may be that a sufficiently weak Feshbach resonance can be found amongst the many resonances of $^{87}$Rb \cite{Marte02}.

An experimental demonstration of collective dissociation thresholds, can however be facilitated in setups where atoms and molecules are optically coupled via a stimulated Raman transition \cite{Wynar2000,Jaksch02,Rom04,Winkler05,Ryu06,Stoferle06}. In this case, $g=\Omega_p\Omega_s/2\Delta_{2p}$ is an effective two-photon Rabi frequency. The one-photon Rabi frequencies $\Omega_{s,p}$ for the pump (inducing a bound-bound transition to an intermediate molecular state) and the Stokes (dissociating the intermediate state via a bound-free transition) lasers, respectively, are products of laser intensities by overlap integrals consisting of electronic transition dipole moments and Franck-Condon factors, and $\Delta_{2p}$ is the two-photon detuning from the intermediate bound state. Thus, the effective coupling strength can be controlled with great precision \cite{Rom04} through the adjustment of laser parameters. Most appealing are systems of molecules in optical lattices \cite{Jaksch02,Rom04,Ryu06,Stoferle06}, providing optical coupling to deeply bound internal molecular states, as well as tight trap frequencies of $\omz/2\pi=10-100$ kHz. Currently, these experiments seek to avoid collective effects in association by operating in the Mott-insulator regime with unit occupation numbers. However, provided that molecular condensates containing roughly $10^2-10^3$ particles per site be formed, their dissociation can demonstrate the expected amplification threshold.  

\begin{figure}
\centering
\includegraphics[scale=0.3,angle=-90]{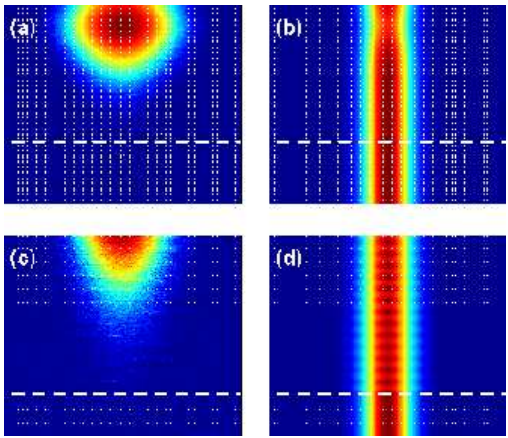}
\caption{(color online) Dynamics of dissociation following a sudden switch in $g$ from $0.9\hbar\omz/\sqrt{n}$ to $\hbar\omz/\sqrt{n}$ (a-b) or in $\omz$ from $1.1g\sqrt{n}/\hbar$ to $g\sqrt{n}/\hbar$ (c-d). Dashed lines mark the time of the switch at $\omz t=4\pi$. Atomic density distributions are shown in (a) and (c) using a logarithmic colormap, whereas the molecular condensate density is plotted in (b) and (d) on a linear colormap. Molecular interaction and effective deuning are $\um=\delta=0$.}
\label{bosdis_fig4}
\end{figure}

Given $10^3$ $^{87}$Rb$_2$ molecules in a $1$ kHz trap, the critical value $g_c=\hbar^2/(m\sqrt{N\lz/\Lambda_3})$ of the interaction strength turns out to be approximately $2\pi\hbar\times 2.2\times10^{-5}$ Hz cm$^{3/2}$. The corresponding average atomic density is of order $10^{13}-10^{14}$ cm$^{-3}$ and $a_s\approx100$ bohr for $^{87}$Rb. Therefore, $Un\approx 75-750$ Hz for a fully dissociated gas, and atomic interactions are initially negligible with respect to the trap and coupling frequencies. The proposed experiment will involve the preparation of a molecular BEC in a trap smaller than the resonance healing length, thus arresting dissociation even in the presence of coupling. A sudden change either in $g$ or in $\omz$ (which in a lattice can be achieved via switching the lattice wavelength or depth) will trigger the stimulated dissociation of the BEC. Both scenarios are depicted in Fig. \ref{bosdis_fig4} for a 1D calculation. In Figs. \ref{bosdis_fig4}a and \ref{bosdis_fig4}b, atomic and molecular densities are shown as a function of time for a $g$-switch experiment. Evidently, dissociation is triggered by the change in the coupling frequency. Similarly, a sudden reduction in trap frequency (Figs. \ref{bosdis_fig4}c,d) can lead to stimulated dissociation of the breathing molecular condensate. When the simulation is carried over the same timescale without changing $g$ or $\omz$, no dissociation is observed. 

In conclusion, progress in tight confinement of molecular quantum gases and control over atom-molecule coupling, brings trap frequencies to the vicinity of atom-molecule conversion frequencies. Molecular BECs can be stabilized against Bose-stimulated dissociation, if confined to a trap smaller than the resonance healing length $\zeta$. Critical values depend on condensate shape and trap geometry. Experimentally, dissociation can be triggered by a sudden change in trap parameters. Such an experiment requires that (a) $\zeta\sim\lz$ to arrest dissociation, (b) $n\lz^3\gg 1$ so that collective effects are significant, and preferably (c) $\xi\gg\zeta$ so that inelastic collisions between atoms do not interfere with association. Future work will explore pattern formation in 2D and 3D tight-trapping experiments. 

This work was supported by grants from the Minerva foundation
for a junior research group and the Israel Science Foundation for a
Center of Excellence (grant No.~8006/03).


\begin{thebibliography}{99}

\bibitem{Drummond98}
P. D. Drummond {\it et al.}, Phys. Rev. Lett. {\bf 81}, 3055 (1998).


\bibitem{Javanainen99}
J. Javanainen and M. Mackie, Phys. Rev. A {\bf 59}, R3186 (1999).

\bibitem{Kostrun99}
J. Javanainen and M. Kostrun, Opt. Exp. {\bf 5}, 188 (1999).

\bibitem{Vardi01}
A. Vardi, V. Yurovsky, and J. R. Anglin, Phys. Rev. A {\bf 64}, 063611 (2001).

\bibitem{Meystre05}
P. Meystre, J. Phys. B: At. Mol. and Opt. Phys. {\bf 38}, S617 (2005).

\bibitem{Rowen05}
E. Rowen {\it et al.}, Phys. Rev. A {\bf72}, 053633 (2005).

\bibitem{YarivWalls}
A. Yariv, {\it Optical Electronics}, 3rd ed. (Holt Rinehart, and Winston, New York, 1985);
D. F. Walls and G. J. Milburn, {\it Quantum Optics}, 1st ed. (Springer-Verlag, Berlin, 1994).

\bibitem{superchem}
D. J. Heinzen {\it et al.},  Phys. Rev. Lett. {\bf 84}, 5029, (2000);
Y. Wu and R. Cote, Phys. Rev. A {\bf 65}, 053603 (2002);
M. W. Jack and H. Pu, {\it ibid.} {\bf 72}, 063625 (2005).

\bibitem{Moore02}
M. G. Moore and A. Vardi,  Phys. Rev. Lett. {\bf 88}, 160402 (2002)

\bibitem{quantpa}
J. J. Hope and M. K. Olsen, Phys. Rev. Lett. {\bf 86}, 3220 (2001);
U. V. Poulsen and K. Molmer, Phys. Rev A {\bf 63}, 023604 (2001);
G. R. Jin, C. K. Kim, and K. Nahm, {\it ibid.} {\bf 72}, 045602 (2005);
P. Naidon and F. Masnou-Seeuws, {\it ibid.}{\bf 73}, 043611 (2006).

\bibitem{Kheruntsyan}
K. V. Kheruntsyan and P. D. Drummond, Phys. Rev A {\bf 66}, 031602(R) (2002);
K. V. Kheruntsyan, {\it ibid.} {\bf 71}, 053609 (2005). 

\bibitem{WallsYurke}
D. F. Walls and C. T. Tindle, J Phys. A: Gen. Phys. {\bf 5}, 534 (1972);
B. Yurke {\it et al.}, Phys. Rev. A {\bf 35}, 3586 (1987).

\bibitem{Greiner03}
M. Greiner, C. A. Regal, and D. S. Jin, Nature {\bf 426}, 537 (2003).

\bibitem{Tiesinga93}
E. Tiesinga, B. J. Verhaar, and H. T. C. Stoof, Phys. Rev A {\bf 47},4114 (1993).

\bibitem{Band95}
Y. B. Band and P. S. Julienne,  Phys. Rev A {\bf  51}, R4317 (1995).

\bibitem{Pethick}
C. J. Pethick and H. Smith, {\it Bose-Einstein condensates in dilute gases}, 
1st ed. (Cambridge University Press, Cambridge, 2002); L. Pitaevskii and S. Stringari, {\it Bose-Einstein condensation}, 1st ed. (Clarendon Press, Oxford, 2003).

\bibitem{Collapse}
H. T. C. Stoof, Phys. Rev. A {\bf 49}, 3824 (1994);
J. M. Gerton, D. Strekalov, I. Prodan, and R. G. Hulet, Nature {\bf 408}, 692 (200);
E. A. Donley {\it et al.}, {\it ibid.} {\bf 412}, 295 (2001).

\bibitem{Attractive}
P. A. Ruprecht {\it et al.}, Phys. Rev. A {\bf 51}, 4704 (1995);
C. C. Bradley {\it et al.}, Phys. Rev. Lett. {\bf 75}, 1687 (1995);
C. C. Bradley {\it et al.}, {\it ibid.} {\bf 78}, 985 (1997).

\bibitem{Santagiustina98}
M. Santagiustina {\it et al.}, Phys. Rev. E {\bf 58}, 3843 (1998).

\bibitem{Gatti97}
A. Gatti {\it et al.}, Phys. Rev. A {\bf 56}, 877 (1997).


\bibitem{Inouye98}
S. Inouye {\it et al.}, Nature (london) {\bf 392} 151 (1998).

\bibitem{Stenger99} 
J. Stenger {\it et al.},   Phys. Rev. Lett {\bf 82}, 2422 (1999).

\bibitem{Timmermans99}
E. Timmermans {\it et al.}, Phys. Rev. Lett {\bf 83}, 2691 (1999); E. Timmermans {\it et al.}, Phys. Rep. {\bf 315}, 199 (1999).

\bibitem{Holland01}
M. Holland, J. Park, and R. Walser, Phys. Rev. Lett {\bf 86}, 1915 (2001).

\bibitem{Marte02}
A. Marte {\it et al.}, Phys. Rev. Lett {\bf 89}, 283202 (2002).

\bibitem{Wynar2000}
R. Wynar {\it et al.}, Science {\bf 287}, 1016 (2000).

\bibitem{Jaksch02}
D. Jaksch {\it et al.}, Phys. Rev. Lett {\bf 89}, 040402 (2002).

\bibitem{Rom04}
T. Rom {\it et al.}, Phys. Rev. Lett {\bf 93}, 073002 (2004).

\bibitem{Winkler05}
K. Winkler {\it et al.}, Phys. Rev. Lett {\bf 95}, 063202 (2005).

\bibitem{Ryu06}
C. Ryu {\it et al.}, cond-mat/0508201.

\bibitem{Stoferle06}
T. St\"oferle {\it et al.}, Phys. Rev. Lett {\bf 96}, 030401 (2006).
\end{thebibliography}
\end{document}